\documentclass[aps,prb,twoside,superscriptaddress,longbibliography,nofootinbib,floatfix, notitlepage,twocolumn]{revtex4-1}

\usepackage{hyperref}
\hypersetup{
  colorlinks = true, 
  linkcolor={blue!50!black}, 
  citecolor=red, 
  urlcolor=blue, 
  linktoc=page,
}

\usepackage[utf8]{inputenc}
\usepackage{color}
\usepackage{amsmath}
\usepackage{braket}
\usepackage{comment}
 \usepackage{amssymb}
 \usepackage{amsmath}
 \usepackage{lipsum, babel}

\usepackage{graphicx}
\usepackage{helvet}
\usepackage{sansmath} 
\usepackage{bm}
\usepackage{appendix}

\usepackage{pgf}
\usepackage{tikz}
\usetikzlibrary{quantikz}
\DeclareGraphicsExtensions{.png,.pdf}

\begin{document}
\title{
How pure can we go with  adiabatic state manipulation?
}

\author{Raul A. Santos},  
\affiliation{Phasecraft Ltd.}

\author{Alex Kamenev}
\affiliation{School of Physics and Astronomy, University of Minnesota,
Minneapolis, Minnesota 55455, USA}
\affiliation{William I. Fine Theoretical Physics Institute, University
of Minnesota, Minneapolis, Minnesota 55455, USA}

\author{Yuval Gefen}
\affiliation{Department of Condensed Matter Physics, The Weizmann Institute
of Science, Rehovot 76100, Israel}

\date{\today}

\begin{abstract}

Dissipative systems with decoherence free subspaces, a.k.a. dark spaces (DSs), can be used to
protect quantum information. At the same time, dissipation is expected to give rise to coherent information degradation outside the DS. Employed to support quantum information platforms, DSs can be
adiabatically modified in a way that resembles adiabatic control of coherent systems. Here we study the slow     evolution of a purely dissipative system with a spectral gap $\gamma$, characterized by a strong
symmetry, under a cyclic protocol with period $T$. 
Non-adiabatic corrections to the state evolution give rise to decoherence: the evolution within the instantaneous DS is described by a time-local effective Liouvillian operator that leads to purity degradation over a period, of order $1/\gamma T$. We obtain a closed form of the latter to order $1/(\gamma T)^2$. Our analysis underlines speed limitations in quantum information processing in the absence of corrective measures. 
\end{abstract}
\maketitle

The density matrix of a quantum state, averaged over stochastic fluctuations reflecting the quantum nature
of the dynamics, evolves according to Lindblad equation. The latter could result from drive-and-
dissipation dynamics \cite{Bardyn_2013} , or from (passive) measurement-induced platforms \cite{Roy2020},
and may be used to steer a system towards a desired target state. When the steady state of the Lindblad
dynamics is multiply degenerate one refers to it as a dark space (DS) that can serve as a basis for quantum computation \cite{Wu2005,Xu2012}. As discussed in \cite{Lieu2020,Santos2020}, in an open
system the nature of the steady states and their ability to serve as generalized quantum bits are determined by the
symmetry of the system, which can appear in a weak or a strong sense \cite{Lieu2020}. While a weak symmetry renders a
state in the DS a classical register, a strong symmetry facilitates the realization of a qubit. Evidently, this requires quantum coherence in the underlying protected subspace.

Dark spaces (also known as decoherence-free spaces) are thus  clear candidates for a quantum information processing platform \cite{Lidar2014}, and indeed have been realized in experiments \cite{Kwiat2000}. 
The steady states of a dissipative evolution have been used to protect Schr\"{o}dinger cat states {\cite{Leghtas2015}, Bell states \cite{Brown_2022} and increasing bit flip time \cite{Berdou2023} on superconducting architectures coupled to external reservoirs.

The questions raised and answered in the present manuscript are two-fold: (i) how to manipulate the DS of a  system to achieve an arbitrary on-demand state within the DS, and (ii) what are the limits on the quality of such manipulations, i.e, the purity of the accessible states.  
The reason why the first question is non-trivial is that by the very definition of a DS it is immune against external signals. We show that state manipulation may be achieved by an adiabatic cyclic rotation of the DS within the larger Hilbert space of the system. Upon such rotation DS states undergo a (in general non-Abelian) Berry rotation\cite{Avron_2012}, which may be engineered to lead to a desired outcome.  We note that adiabatic manipulations have been reported for single and multi-particle \cite{Toyoda2013,Leroux2018,Sugawa_2021,Wang2022,FerrerGarcia2022} platforms. 

The second question concern limitations on the precision of such (nearly) adiabatic manipulations. Within unitary quantum dynamics an adiabatic evolution may be contaminated by Landau-Zener transitions into undesired states. The probability of such events is known  to be exponentially small in the adiabatic parameter \cite{landau2013quantum,Zener_1932}, i.e. product of the spectral gap, $\gamma$, and a characteristic time scale, $T$, of the adiabatic signal. What is the analog of such a Landau-Zener escape for the dissipative Lindbladian evolution? Does it, e.g., suppress the purity of the states, and how does such a suppression scale with the adiabatic parameter? We show here that the Berry phase rotation is always associated with the purity degradation. Moreover, the latter scales only {\em algebraically} (as the minus first power) with the adiabatic parameter, $\gamma T$.

The protocols studied in this work involve closed cycle rotations of the DS, by time-dependent Linbladians, $\mathcal{L}_t$, with  $\mathcal{L}_0 = \mathcal{L}_T$, where $T$ is the period of the DS rotation. We focus on purely dissipative systems with the spectral gap $\sim \gamma$. The existence of a DS is guaranteed by an instantaneous strong symmetry.  The cyclic variation of the DS is intimately related to the notion of Abelian and non-Abelian geometric
phases\cite{Zanardi1999,Zhang2023}, and has been extended beyond the adiabatic regime \cite{Xu2012}.   Here we find that non-adiabatic corrections to the DS evolution give
rise to the purity degradation of the order $1/(\gamma T)$. To derive this result we show that the slow evolution within the DS, to the
order $1/(\gamma T)$, may be described by an effective Markovian evolution and derive the corresponding Linbladian operator.  

We are interested in studying a Markovian evolution of the reduced density matrix, described the Lindblad equation
\begin{equation}\label{eq:evol}
    \frac{d\varrho}{d t}=\mathcal{L}_t[\varrho]=\gamma\left(L_t \varrho L^\dagger_t -\frac{1}{2}\{L^\dagger_t L_t,\varrho\}\right),
\end{equation}
with the quantum jump operator $L_t$, which is slowly varying in time. 
Here the rate $\gamma$ fixes the size the spectral gap of the Lindbladian. Such an evolution can be generated by, e.g., a protocol of weak measurements where the
direction of measurement is changed continuously, in the
limit of infinitesimal steps \cite{Snizhko2019prl,Snizhko2019,Snizhko_2021,Snizhko_2021b}.  To be specific, we assume that in the ``laboratory frame'' the jump operator is given by a time-dependent unitary rotation of a certain fixed 
``rotating frame" jump operator $L$, i.e. 
\begin{equation}
    \label{Eq:rotating-frame}
L_t = \mathcal{U}^\dagger_t L\, \mathcal{U}_t,     \end{equation}
where $\mathcal{U}_t$ is periodic in time with the period $T$, $\mathcal{U}_{t+T}=\mathcal{U}_t$, and $\mathcal{U}_0=\mathcal{U}_T=1$.
Moreover, we assume that there is a $d$-dimensional dark space in the rotating frame, spanned by $d$ states $|m\rangle$ that are annihilated by $L$,  
\begin{equation}
    L|m\rangle = 0 = \langle m|L^\dagger,\quad m\in (1,\dots,d).
\end{equation}
i.e the system possesses a strong symmetry \cite{Lieu2020}. 

In the laboratory frame the dark space is slowly rotated and is instantaneously spanned by the vectors $|m_t\rangle=\mathcal{U}^\dagger_t |m\rangle$. 
If not for this slow rotation, the system would end up in one of the states (pure or mixed) within the dark space at long times, where the evolution would stop completely. Because of the slow rotation, the system has to constantly catch up to the instantaneous dark space,  undergoing a fast dissipative evolution towards it. As a result, the state of the system within the dark space keeps slowly evolving at all times. 

Our goal is to study such a residual slow evolution of the $d\times d$ dimensional projection of the full density matrix onto the rotated dark space. 
To this end we first pass to the rotated frame, where the rotated density matrix
$\rho(t) =  \mathcal{U}_t \varrho(t) \mathcal{U}^\dagger_t$ obeys the following evolution equation
\begin{equation}\label{eq:evol_rot}
    \frac{d\rho}{d\tau}=-i\frac{1}{\gamma T}[{H}_\tau,\rho]+\left(L\rho L^\dagger -\frac{1}{2}\{L^\dagger L,\rho\}\right),
\end{equation}
where we switched to the dimensionless time $\tau = \gamma t\in [0,\gamma T]$, where $\gamma T\gg 1$. Here the effective adiabatic Hermitian Hamiltonian 
\begin{equation}
                \label{eq:eff-ham}
    H_\tau = i\gamma T \, (\partial_\tau \mathcal{U})\mathcal{U}^\dagger_\tau 
\end{equation}
is generated. Note that besides the effective unitary evolution in Eq. (\ref{eq:evol_rot}), our dynamics is controlled by a Lindbladian, that we denote by $\mathcal{L}[\rho]$. It provides the aforementioned fast evolution towards the DS. The 
effective Hamiltonian (\ref{eq:eff-ham}) provides a small, $1/(\gamma T)\ll 1$, correction to this fast dissipative evolution. Since the system fast evolves towards the DS, we now introduce DS projector 
 \begin{equation}\label{eq:projector}
     P_0 = \sum_{m=1}^d|m\rangle\langle m|, 
 \end{equation}
and the projected $d\times d$ effective density matrix within the DS, defined as $\rho_0(\tau) = P_0 \rho(\tau) P_0$. 

Projecting Eq.~(\ref{eq:evol_rot}) (as discussed below) onto the instantaneous  (continuously rotating) DS, we find remarkably that up to order $1/(\gamma T)^2$, the evolution of the projected density matrix is Markovian and its averaged time evolution is given by an effective Lindbladian. We notice that 
the latter statement is rather non-trivial. Even more so, we have found an explicit analytic recipe to generate effective jump operators underlying the evolution within the DS. Our findings are summarized (see the Supplemental Material (SM) for a complete derivation) by the evolution equation for the DS projected density matrix $\rho_0(\tau)$ 
\begin{align}\label{eq:main_res}\nonumber
    \frac{d \rho_0}{d \tau} &=-\frac{i}{\gamma T}\,[H^0_\tau,\rho_0] +\frac{1}{(\gamma T)^2}\left(\ell_\tau\rho_0\ell^{\dagger}_\tau-\frac{1}{2}\{\ell^{\dagger}_\tau\ell_\tau,\rho_0\}\right)\\
    &+O(1/\gamma^3 T^3),
\end{align}
with $H^0_\tau:=P_0H_\tau P_0$, and $\ell_\tau:=2P_0 L (L^\dagger L)^{-1}(1-P_0)H_\tau P_0$.
Here $(L^\dagger L)^{-1}$ is the Moore-Penrose pseudoinverse of $L^\dagger L$ \cite{ben2003generalized}.

Equation (\ref{eq:main_res}) for the evolution of the projected density matrix constitutes the main result of this letter. It states that to the leading order in $1/(\gamma T)$ such an evolution is unitary (see also Ref. \onlinecite{Avron_2012}). In other words, a state of the system within the DS undergoes a unitary evolution, acquiring a non-Abelian Berry phase \cite{Wilczek1984}, and preserving the state's purity. Yet, since the dissipative evolution towards the instantaneous DS is involved, there must be  some purity degradation. The latter is described by $1/(\gamma T)^2$ Lindbladian term in Eq.~(\ref{eq:main_res}).

Taking an initial state $\varrho(0)$ (in the laboratory frame) fully within the DS, at the end of the cycle (and up to first order in $1/\gamma T$) the density matrix is, 
\begin{align}\label{eq:final_dm}
    &\varrho({T})= U(\gamma T,0)^\dagger\varrho(0)U(\gamma T,0)\\\nonumber  &+\frac{U(\gamma T,0)^\dagger}{(\gamma T)^2}\left[\int\limits_0^{\gamma T}\! d\tau \left(l_\tau\varrho(0) l^{\dagger}_\tau-\frac{1}{2}\{l^{\dagger}_\tau l_\tau,\varrho(0)\}\right)\right]U({\gamma T},0),
    \end{align}
where the non-Abelian Berry phase 
\begin{equation}
    \label{Eq:non-abelian-Berry}
U(\tau,0)=\mathcal{T}\exp\left\{\frac{i}{\gamma T}\int\limits_0^{\tau}ds\, H_s^0\right\}    
\end{equation} 
is given by the time ordered exponential of the effective Hamiltonian $H^0_\tau$ \cite{Wilczek1984}, and
$ l_\tau=U(\tau,0)\ell_\tau U(\tau,0)^\dagger$.
 Up to the order $1/\gamma T$ the purity degradation of the initial density matrix of purity ${\rm Tr}(\varrho^2(0))=\Gamma_0$ is given by
\begin{equation}
    {\rm Tr}(\varrho^2({T}))=\Gamma_0 - \frac{2}{(\gamma T)^2}\int\limits_{0}^{\gamma T}\! ds\, {\rm Tr}\left[[\varrho(0), l^\dagger_{s}] l_{s}\varrho(0)\right].
\end{equation}
 This shows that the purity degradation is of the order $(\gamma T)^{-1}$. At the same time the spectral weight escape out of the DS is exponentially small in $\gamma T\gg 1$, similarly to the coherent Landau-Zener scenario. Therefore the loss of quantum information due to purity degradation within the DS is parametrically more significant than due to its leakage out of the DS. This is the main message of the present letter.

Here we summarize only the main idea of the derivation of this result and refer to the SM for further details. One starts with Eq. (\ref{eq:evol_rot}) in the rotated frame. Using the projector $P_0$ and its complement $1-P_0$, one projects this equation into four different subspaces, acting on the left (right) with $P_0$ or $1-P_0$. This generates four linear coupled differential equations for the derivatives of $\rho_0:=P_0\rho P_0$, $\rho_{10}:=(1-P_0)\rho P_0$, $\rho_{01}:=P_0\rho(1-P_0)$ and $\rho_{11}:=(1-P_0)\rho(1-P_0)$.
Next, one would like to eliminate the fast relaxation dynamics and describe the evolution within the slow submanifold characterized by $\rho_0$. One can achieve this by first solving the equation for $\rho_{11}$, in terms of the off-diagonal operators $\rho_{10}$ and $\rho_{01}$. As the equations for $\rho_{10}$ and $\rho_{01}$ also depend on $\rho_{11}$, plugging back its expression in terms of $\rho_{10}(\rho_{01})$ into that equation generates an integral equation that can be solved through Picard iteration. This procedure naturally produces a series in $\epsilon=(\gamma T)^{-1}$.
To first order in $\epsilon$, the solution of $\rho_{10}(\rho_{01})$ depends only on $\rho_0$. This allows one to write a closed expression for the evolution of $\rho_0$, that has a Lindblad form given by Eq. (\ref{eq:main_res}).

 To illustrate the results, let us 
 consider the evolution of a spin 3/2 system, similar to the one used by Wilczek and Zee \cite{Wilczek1984,Zee1988} in the context of the non-Abelian berry phases. The system has a 4-dimensional Hilbert space, and a 2-dimensional dark space within it. Unlike 
 Refs.~[\onlinecite{Wilczek1984,Zee1988}], we consider a pure Lindbladian evolution with a time dependent jump operator:
\begin{equation}
\label{eq:jump_ex}
     L_t =\sqrt{\gamma} \, e^{-i\phi S_{z}}e^{-i\theta S_{y}}S_{x}\left(S_{z}^{2}-\frac{1}{4}\right)e^{i\theta S_{y}}e^{i\phi S_{z}}, 
 \end{equation}
 where $S_a, a={x,y,z}$ are $4\times 4$ spin 3/2 operators, and $(\theta,\phi)$ are smooth functions of time such that $(\theta,\phi)(T)=(\theta,\phi)(0)+2\pi (m_\theta,m_\phi)$ with $m_\theta,m_\phi$ integers. 
 The dark space of this system is spanned by the rotated states $\left|\pm(t)\right\rangle =e^{-i\phi S_{z}}e^{-i\theta S_{y}}\left|\pm\frac{1}{2}\right\rangle$, where $S_z\left|\pm\frac{1}{2}\right\rangle = \pm\frac{1}{2} \left|\pm\frac{1}{2}\right\rangle$,   that satisfy $L_t\left|\pm(t)\right\rangle =0$.

This dissipative evolution can be engineered by coupling a spin 3/2 state to an optical cavity with strong symmetry in the spin 1/2 subspace. The latter could be achieved by driving with a laser the states of spin $m=\pm 3/2$ in resonance with the cavity. By detuning the drive off-resonance a time dependent rotation could be achieved. A protocol of this nature has been proposed and analyzed in Ref. \onlinecite{young2024}, where collective spin excitations are used to generate spin-squeezed states through Berry phase engineering.
 
 The effective Hamiltonian is $H_\tau=i\gamma T\,(\partial_\tau \mathcal{U})\mathcal{U}^\dagger$, where $\mathcal{U}_\tau=e^{i\theta S_{y}}e^{i\phi S_{z}}$. Explicitly, this corresponds to
 \begin{eqnarray}
     H = -\gamma T\left(\frac{d\theta}{d\tau}S_y+(\cos\theta S_z-\sin\theta S_x)\frac{d\phi}{d\tau}\right).
 \end{eqnarray}
 Its DS projection $H_\tau^{0}=P_{0}H_\tau P_{0}$ with  $P_0=\frac{1}{2}\left(\frac{9}{4}-S_z^2\right)$ is
 \begin{equation}
     H_\tau^0=-\theta'\sigma_{y}-\phi'\left(\frac{1}{2}\cos\theta\,\sigma_{z}-\sin\theta\,\sigma_{x}\right),
 \end{equation}
 where $\sigma_{x,y,z}$ are the Pauli matrices in the $|\pm\frac{1}{2}\rangle$ subspace and $(\theta',\phi')=\frac{d}{ds}(\theta,\phi)$, $s=t/T$.
 
The effective quantum jump operator in this subspace is
 \begin{align}\nonumber \ell_\tau=P_0L\int\limits_{0}^{\tau}ds\,e^{\frac{1}{2}L^{\dagger}L(s-\tau)}(1-P_0)H_sP_{0}=a_\tau\boldsymbol{1}+ib_\tau\sigma_{z}
 \end{align}
 with 
 \begin{align}
     a_\tau=\frac{3}{2}\int\limits_{0}^{\tau}ds\, e^{\frac{3}{2}(s-\tau)}\phi'\sin\theta; \qquad
     b_\tau=\frac{3}{2}\int\limits_{0}^{\tau}ds\, e^{\frac{3}{2}(s-\tau)}\theta'.
 \end{align}
The evolution in the $|\pm \frac{1}{2}\rangle$ subspace for this system is given by the effective Lindbladian
\begin{align}
    \mathcal{L}_\tau^{\rm DFS}[\rho_0]&=-\frac{i}{\gamma T}[H^0,\rho_0]\\
    &+\frac{1}{(\gamma T)^2}\left(ia_\tau b_\tau[\sigma_z,\rho_0]+b_\tau^{2}(\sigma_z\rho_0\sigma_z-\rho_0)\right), \nonumber 
\end{align}
Starting from a pure density matrix in the $|\pm \frac{1}{2}\rangle$ DS, parameterised by $\rho_0(0)=\frac{1}{2}(1+\vec{n}_0\cdot\vec{\sigma})$, with $\vec{n}_0=(n^x_0,n^y_0,n^z_0)$ a unit vector ($\vec n_0^2=1$), we find that, to the leading order in $1/(\gamma T)$, the DS density matrix experiences the non-Abelian rotation, parameterized by $n^a_\tau=\frac{1}{2}{\rm Tr}\left\{\sigma_a U(\tau,0)^\dagger(\vec{n}_{0}\cdot\vec{\sigma}) U(\tau,0)\right\}$ and $U(\tau,0)$ -- the  Berry rotation, Eq.~(\ref{Eq:non-abelian-Berry}). 
However, to the next order the purity (and thus the length of the $\vec n_T$ vector) is suppressed by the factor $\Gamma_T\equiv {\rm Tr}(\rho_T^2)$, given by 
\begin{align}\label{eq:pur}
    \Gamma_T=1-\frac{2}{(\gamma T)^2}\int\limits_0^{\gamma T}d\tau\,  b_\tau^2\, \left((n^x_\tau)^2+(n^y_\tau)^2\right).
\end{align}
In the simplest case of $\theta=2\pi\frac{t}{T}$, this leads to
\begin{align}\label{ref:pred}
    \Gamma_T = 1 - 4\pi^2\, \frac {1+(n^y_0)^2}{\gamma T}+O\left(\frac{1}{(\gamma T)^2}\right).
\end{align}

\begin{figure}[ht!]
    \centering
\includegraphics[scale=0.8]{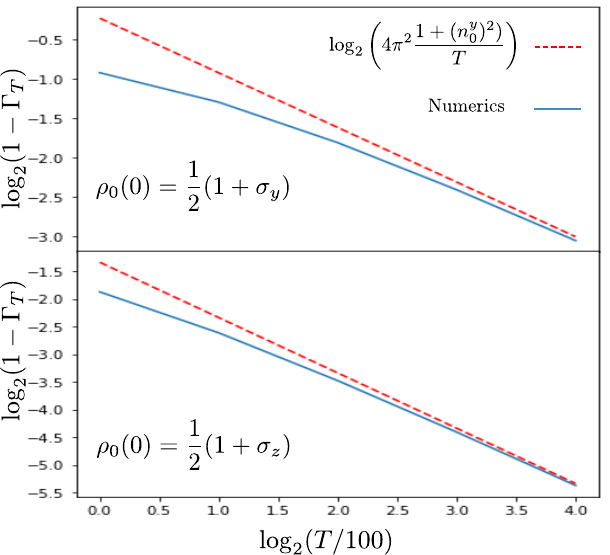}
    \caption{Purity degradation as defined by $1-\Gamma_T$ in the spin $\frac{3}{2}$ system with jump operator Eq. (\ref{eq:jump_ex}). We plot this as a function of the period $T$ (we have set the rate $\gamma=1$). The numerical results approach the prediction of Eq. (\ref{eq:pur}) as the period increases. Here the initial state is $\rho_0(0)=\frac{1}{2}(1+\sigma_y)$ (top panel) which corresponds to $n^y_0=1$ and $\rho_0(0)=\frac{1}{2}(1+\sigma_z)$ (bottom panel), i.e $n_0^y=0$. }
    \label{fig:rho_ny}
\end{figure}

Notice that in this latter simple scenario there is no Berry phase accumulated over the cycle. Yet, the purity is degraded. 
To benchmark out theoretical findings, we performed numerical simulations of the model described above, with two different initial states, shown in Fig (\ref{fig:rho_ny}) top and bottom. In particular, we computed the purity degradation at the end of the period, as a function of the period $T$. We find excellent agreement with our theory prediction.

In the adiabatic limit, it is known \cite{Wilczek1984} that the non-Abelian Berry phase, acquired over a closed cycle, is independent of the basis chosen to span the Hilbert space. This occurs because different basis choices correspond to gauge transformations of the non-Abelian Berry curvature, whose action cancels on a closed loop. Since the definitions of the projected Hamiltonian,  $H_{0}$, and the effective jump operator, $\ell_\tau$, rely on a specific basis within the DS, they both 
are sensitive to the change of gauge. One may wonder thus if, e.g., the purity degradation  is a gauge invariant quantity. 
In the SM we show that this is indeed the case. 

The main message of this letter may be summarized as a ``glass half filled''. Indeed, one can achieve an arbitrary non-Abelian rotation within the DS by performing its proper rotation in the larger Hilbert space. This is certainly a valuable asset for quantum manipulations. 
On the flip side, the non-adiabatic, purity-degrading effects are only algebraically (not exponentially) suppressed by the adiabatic parameter. This imposes rather stringent requirements on the rates of such operations.

We are grateful to Yosi Avron and Victor Albert for illuminating discussions.  
The work was supported by the NSF Grants No. DMR-2037654 and DMR-2338819. Y.G. was
supported by the DFG grant EG 96/13-1 and NSF-BSF 2023666, and is the incumbent of InfoSys Chair (at IISc).

\bibliography{bibliography}

\pagebreak

\onecolumngrid

\appendix

\section*{Supplemental material:
How pure can we go with  adiabatic state manipulation?}

\subsection{Solution to the evolution equation}\label{app}
From the main text, we want to solve
\begin{align}
\frac{d}{d\tau}\rho & =-i\epsilon[H_{\tau},\rho]+\left(L\rho L^{\dagger}-\frac{1}{2}\{L^{\dagger}L,\rho\}\right),\label{eq:evolution}
\end{align}
with the followings assumptions:
\begin{enumerate}
\item  $\rho(0)\in$ DS, i.e the initial density matrix is completely within
the dark space. We call the projector into the DS $P_{0}$
\item  $(1-P_{0})L(1-P_{0})=0$, which correspond to the condition that
the DS of the evolution is fully characterized by $P_{0}$
\item $\|\frac{dH_\tau}{d\tau}\|=O(\epsilon)$, i.e the evolution is adiabatic.
\end{enumerate}

We use a straightforward approach to solve this problem. We first
decompose the evolution Eq. (\ref{eq:evolution}) into four possible orthogonal
spaces, using the projector onto de DS $P_{0}$. This leads to four
coupled linear differential equations for the projections of the density
matrix into the four spaces. We then solve these equations as an expansion in the parameter $\epsilon$ as follows.

Using $P_{0}$, we can decompose any operator into four projections
\begin{align*}
A_{00} & :=P_{0}AP_{0},\\
A_{10} & :=(1-P_{0})AP_{0},\\
A_{01} & :=P_{0}A(1-P_{0}),\\
A_{11} & :=(1-P_{0})A(1-P_{0}).
\end{align*}
Then Eq. \ref{eq:evolution} becomes the set of four equations

\begin{align}
\frac{d}{d\tau}\rho_{00} & =-i\epsilon\left([H_{\tau}^{00},\rho_{00}]+H_{\tau}^{01}\rho_{10}-\rho_{01}H_{\tau}^{10}\right)+L_{01}\rho_{11}L_{10}^{\dagger},\nonumber \\
\frac{d}{d\tau}\rho_{01} & =-i\epsilon[H_{\tau}^{00}\rho_{01}+H_{\tau}^{01}\rho_{11}-\rho_{00}H_{\tau}^{01}-\rho_{01}H_{\tau}^{11}]+L_{01}\rho_{11}L_{11}^{\dagger}-\frac{1}{2}\rho_{01}(L^{\dagger}L)_{11},\nonumber \\
\frac{d}{d\tau}\rho_{10} & =-i\epsilon[H_{\tau}^{10}\rho_{00}+H_{\tau}^{11}\rho_{10}-\rho_{10}H_{\tau}^{00}-\rho_{11}H_{\tau}^{10}]+L_{11}\rho_{11}L_{10}^{\dagger}-\frac{1}{2}(L^{\dagger}L)_{11}\rho_{10},\label{eq:set_proj}\\
\frac{d}{d\tau}\rho_{11} & =-i\epsilon\left([H_{\tau}^{11},\rho_{11}]+H_{\tau}^{10}\rho_{01}-\rho_{10}H_{\tau}^{01}\right)+L_{11}\rho_{11}L_{11}^{\dagger}-\frac{1}{2}\{(L^{\dagger}L)_{11},\rho_{11}\}.\nonumber 
\end{align}

Introducing $U(\tau,0):=\mathcal{T}e^{i\int_{0}^{\tau}\epsilon H_{s}^{00}ds}$,
$V(\tau,0):=\mathcal{T}e^{\int_{0}^{\tau}\left(i\epsilon H_{s}^{11}+\frac{1}{2}(L^{\dagger}L)_{11}\right)ds}$,
we define modified projected density matrices
\begin{align*}
\tilde{\rho}_{00} & :=U(\tau,0)\rho_{00}U^{\dagger}(\tau,0),\quad\tilde{\rho}_{01}:=U(\tau,0)\rho_{01}V^{\dagger}(\tau,0),\\
\tilde{\rho}_{10} & :=V(\tau,0)\rho_{10}U^{\dagger}(\tau,0),\quad\tilde{\rho}_{11}:=V^{\dagger}(\tau,0)\rho_{11}V^{\dagger}(\tau,0),
\end{align*}
and projected Hamiltonians
\begin{align*}
\mathcal{H}_{L}^{01} & :=U(\tau,0)H_{\tau}^{01}V^{-1}(\tau,0),\quad\mathcal{H}_{L}^{10}:=V(\tau,0)H_{\tau}^{10}U^{\dagger}(\tau,0),\\
\mathcal{H}_{R}^{01} & :=U(\tau,0)H_{\tau}^{01}V^{\dagger}(\tau,0),\quad\mathcal{H}_{R}^{10}:=V^{\dagger-1}H_{\tau}^{10}U^{\dagger}(\tau,0),
\end{align*}
together with the operators
\begin{align*}
Q & :=V(\tau,0)L_{11}V^{-1}(\tau,0),\quad Q^{\dagger}:=V^{\dagger-1}(\tau,0)L_{11}^{\dagger}V^{\dagger}(\tau,0),\\
S & :=U(\tau,0)L_{01}V^{-1}(\tau,0),\quad S^{\dagger}:=V^{\dagger-1}(\tau,0)L_{10}^{\dagger}U^{\dagger}(\tau,0).
\end{align*}
With these definitions, the set of evolution eqs (\ref{eq:set_proj})
becomes
\begin{align*}
\frac{d}{d\tau}\left(\tilde{\rho}_{00}\right) & =-i\epsilon\left(\mathcal{H}_{L}^{01}\tilde{\rho}_{10}-\tilde{\rho}_{01}\mathcal{H}_{R}^{10}\right)+S\tilde{\rho}_{11}S^{\dagger},\\
\frac{d}{d\tau}\left(\tilde{\rho}_{01}\right) & =-i\epsilon[\mathcal{H}_{L}^{01}\tilde{\rho}_{11}-\tilde{\rho}_{00}\mathcal{H}_{R}^{01}]+S\tilde{\rho}_{11}Q^{\dagger},\\
\frac{d}{d\tau}\left(\tilde{\rho}_{10}\right) & =-i\epsilon[\mathcal{H}_{L}^{10}\tilde{\rho}_{00}-\tilde{\rho}_{11}\mathcal{H}_{R}^{10}]+Q\tilde{\rho}_{11}S^{\dagger},\\
\frac{d}{d\tau}\left(\tilde{\rho}_{11}\right) & -Q\tilde{\rho}_{11}Q^{\dagger}=-i\epsilon\left(\mathcal{H}_{L}^{10}\tilde{\rho}_{01}-\tilde{\rho}_{10}\mathcal{H}_{R}^{01}\right).
\end{align*}
Assumption 2 implies that $Q=Q^{\dagger}=0$, simplifying the previous equations
to, 
\begin{align*}
\frac{d}{d\tau}\left(\tilde{\rho}_{00}\right) & =-i\epsilon\left(\mathcal{H}_{L}^{01}\tilde{\rho}_{10}-\tilde{\rho}_{01}\mathcal{H}_{R}^{10}\right)+S\tilde{\rho}_{11}S^{\dagger},\\
\frac{d}{d\tau}\left(\tilde{\rho}_{01}\right) & =-i\epsilon[\mathcal{H}_{L}^{01}\tilde{\rho}_{11}-\tilde{\rho}_{00}\mathcal{H}_{R}^{01}],\\
\frac{d}{d\tau}\left(\tilde{\rho}_{10}\right) & =-i\epsilon[\mathcal{H}_{L}^{10}\tilde{\rho}_{00}-\tilde{\rho}_{11}\mathcal{H}_{R}^{10}],\\
\frac{d}{d\tau}\left(\tilde{\rho}_{11}\right) & =-i\epsilon\left(\mathcal{H}_{L}^{10}\tilde{\rho}_{01}-\tilde{\rho}_{10}\mathcal{H}_{R}^{01}\right).
\end{align*}
The homogeneous solutions to these equations are $\tilde{\rho}_{00}=C_{1}\rightarrow\rho_{00}^{h}(\tau)=U^{\dagger}(\tau,0)\rho(0)U(\tau,0)$
and $\tilde{\rho}_{10}=\tilde{\rho}_{01}=\tilde{\rho}_{11}=0$, where
we have used the initial conditions according to assumption 1. 

Up to now, we have not performed any approximation. To find the effective
evolution equation for $\tilde{\rho}_{00}$, we follow the simple
strategy
\begin{enumerate}
\item Solve the equation for $\tilde{\rho}_{11}$. This gives us 
\[
\tilde{\rho}_{11}=-i\epsilon\int_{0}^{\tau}\left(\mathcal{H}_{L}^{10}\tilde{\rho}_{01}-\tilde{\rho}_{10}\mathcal{H}_{R}^{01}\right)ds.
\]
\item Replace $\tilde{\rho}_{11}$ in the equations for $\tilde{\rho}_{10}$ and $\tilde{\rho}_{01}$
and solve these equations keeping just contributions up to order $\epsilon^{2}$.
The first step leads us to
\begin{align*}
\frac{d}{d\tau}\left(\tilde{\rho}_{01}\right) & =i\epsilon[\tilde{\rho}_{00}\mathcal{H}_{R}^{01}+i\epsilon\mathcal{H}_{L}^{01}\int_{0}^{\tau}\left(\mathcal{H}_{L}^{10}\tilde{\rho}_{01}-\tilde{\rho}_{10}\mathcal{H}_{R}^{01}\right)ds]\\
\frac{d}{d\tau}\left(\tilde{\rho}_{10}\right) & =-i\epsilon[\mathcal{H}_{L}^{10}\tilde{\rho}_{00}+i\epsilon\int_{0}^{\tau}\left(\mathcal{H}_{L}^{10}\tilde{\rho}_{01}-\tilde{\rho}_{10}\mathcal{H}_{R}^{01}\right)ds\mathcal{H}_{R}^{10}]
\end{align*}
We can formally solve these equations using Picard iteration, e.g.
we write the eq for $\tilde{\rho}_{01}$ as
\[
\tilde{\rho}_{01}=i\epsilon[\int_{0}^{\tau}\tilde{\rho}_{00}\mathcal{H}_{R}^{01}+i\epsilon\int_{0}^{\tau}\mathcal{H}_{L}^{01}\int_{0}^{\tau}\left(\mathcal{H}_{L}^{10}\tilde{\rho}_{01}-\tilde{\rho}_{10}\mathcal{H}_{R}^{01}\right)ds]
\]
and replace it iteratively on itself, generating a series expansion
in $\epsilon$. We have
\begin{align*}
\tilde{\rho}_{01}=  i\epsilon\int_{0}^{\tau}\tilde{\rho}_{00}\mathcal{H}_{R}^{01}ds+O(\epsilon^{3})\quad \mbox{and}\quad
\tilde{\rho}_{10}=-i\epsilon\int_{0}^{\tau}\mathcal{H}_{L}^{10}\tilde{\rho}_{00}ds+O(\epsilon^{3}).
\end{align*}

\item Use these solutions to find a closed form of the equation for $\tilde{\rho}_{00}$. This leads to
\begin{align}\label{eq:rho00}
&\frac{d}{d\tau}\left(\tilde{\rho}_{00}\right)  =-i\epsilon\left(\mathcal{H}_{L}^{01}\tilde{\rho}_{10}-\tilde{\rho}_{01}\mathcal{H}_{R}^{10}\right)+S\tilde{\rho}_{11}S^{\dagger}\\\nonumber 
 & =-\epsilon^{2}\left(\mathcal{H}_{L}^{01}\left\{ \int_{0}^{\tau}\mathcal{H}_{L}^{10}\tilde{\rho}_{00}ds\right\} +\left[\int_{0}^{\tau}\tilde{\rho}_{00}\mathcal{H}_{R}^{01}ds\right]\mathcal{H}_{R}^{10}\right)
 +\epsilon^2 S\left[\int_{0}^{\tau}\!\left[\mathcal{H}_{L}^{10}\left[\int_{0}^{\tau_1}\!\!\tilde{\rho}_{00}\mathcal{H}_{R}^{01}ds\right]+\left\{ \int_{0}^{\tau_1}\!\!\mathcal{H}_{L}^{10}\tilde{\rho}_{00}ds\right\} \mathcal{H}_{R}^{01}\right]d\tau_1\right]S^{\dagger}\\\nonumber &+O(\epsilon^{4})
\end{align}
\end{enumerate}
Now, let's analyse closely the contributions to the previous
equation. Starting from $\mathcal{I}_1:=\int_{0}^{\tau}\mathcal{H}_{L}^{10}\tilde{\rho}_{00}ds$, we have
\begin{align*}
&\mathcal{I}_1  =\int_{0}^{\tau}V(s,0)H_{s}^{10}\rho_{00}U^{\dagger}(s,0)ds =\int_{0}^{\tau}\mathcal{T}e^{\int_{0}^{s}\left(i\epsilon H_{r}^{11}+\frac{1}{2}(L^{\dagger}L)_{11}\right)dr}H_{s}^{10}\rho_{00}U^{\dagger}(s,0)ds\\
 & =\int_{0}^{\tau}e^{\frac{s}{2}(L^{\dagger}L)_{11}}\mathcal{T}e^{\int_{0}^{s}i\epsilon\tilde{H}_{r}^{11}dr}H_{s}^{10}\rho_{00}U^{\dagger}(s,0)ds
\end{align*}
where we have used that the time evolution operator $\mathcal{T}e^{\int_{0}^{s}\left(i\epsilon H_{r}^{11}+\frac{1}{2}(L^{\dagger}L)_{11}\right)dr}$ can be written in the interaction picture as 
\begin{align}
\mathcal{T}e^{\int_{0}^{s}\left(i\epsilon H_{r}^{11}+\frac{1}{2}(L^{\dagger}L)_{11}\right)dr}=e^{\frac{s}{2}(L^{\dagger}L)_{11}}\mathcal{T}e^{\int_{0}^{s}i\epsilon \tilde{H}_{r}^{11}dr}
\end{align}
where $\tilde{H}_{s}^{11}:=e^{\frac{s}{2}(L^{\dagger}L)_{11}}{H}_{s}^{11}e^{-\frac{s}{2}(L^{\dagger}L)_{11}}$ is the Hamiltonian ${H}_{r}^{11}$ transformed by a time dependent similarity transformation. Writing the identity $e^{\frac{s}{2}(L^\dagger L)_{11}}=\frac{d}{ds}\left[2(L^{\dagger}L)_{11}^{-1}e^{\frac{s}{2}(L^{\dagger}L)_{11}}\right]$ where the inverse $((L^\dagger L)_{11})^{-1}$ is well defined as the projection eliminates the zero modes of  $(L^\dagger L)_{11}$, we can integrate by parts to find
\begin{align}\nonumber
&\mathcal{I}_1 =\left[2(L^{\dagger}L)_{11}^{-1}e^{\frac{s}{2}(L^{\dagger}L)_{11}}\mathcal{T}e^{\int_{0}^{s}i\epsilon\tilde{H}_{r}^{11}dr}H_{s}^{10}\rho_{00}U^{\dagger}(s,0)\right]_{0}^{\tau}
-\int_{0}^{\tau}2(L^{\dagger}L)_{11}^{-1}e^{\frac{s}{2}(L^{\dagger}L)_{11}}\frac{d}{ds}\left[\mathcal{T}e^{\int_{0}^{s}i\epsilon\tilde{H}_{r}^{11}dr}H_{s}^{10}\rho_{00}U^{\dagger}(s,0)\right]ds\\
 & =\left[2(L^{\dagger}L)_{11}^{-1}e^{\frac{s}{2}(L^{\dagger}L)_{11}}\mathcal{T}e^{\int_{0}^{s}i\epsilon\tilde{H}_{r}^{11}dr}H_{s}^{10}\rho_{00}U^{\dagger}(s,0)\right]_{0}^{\tau}+O(\epsilon)\nonumber 
\end{align}
where we have used that the derivative of each term in the last expression
above is of order $\epsilon$, as it follows from the previous equations and assumption 3. The term evaluated at $s=0$ that comes from the lower limit of the boundary contribution of the integration by parts is exponentially small compared to the one evaluated at $s=\tau$. Keeping just the term from the upper limit we have
\begin{align*}
\mathcal{H}_{L}^{01}\left\{ \int_{0}^{\tau}\mathcal{H}_{L}^{10}\tilde{\rho}_{00}ds\right\}
&=U(\tau,0)H_{\tau}^{01}2(L^{\dagger}L)_{11}^{-1}H_{\tau}^{10}\rho_{00}U^{\dagger}(\tau,0)+O(\epsilon)
\end{align*}
similarly for $\left[\int_{0}^{\tau}\tilde{\rho}_{00}\mathcal{H}_{R}^{01}ds\right]\mathcal{H}_{R}^{10}$
we find
\begin{align*}
\left[\int_{0}^{\tau}\tilde{\rho}_{00}\mathcal{H}_{R}^{01}ds\right]\mathcal{H}_{R}^{10}
 &=U(\tau,0)\rho_{00}H_{\tau}^{01}2(L^{\dagger}L)^{-1}H_{\tau}^{10}U^{\dagger}(\tau,0)+O(\epsilon)
\end{align*}

Using the same approach on $\mathcal{I}:=S\left(\int_{0}^{\tau}\left\{ \int_{0}^{\tau'}\mathcal{H}_{L}^{10}\tilde{\rho}_{00}ds\right\} \mathcal{H}_{R}^{01}d\tau'+\int_{0}^{\tau}\mathcal{H}_{L}^{10}\left[\int_{0}^{\tau'}\tilde{\rho}_{00}\mathcal{H}_{R}^{01}ds\right]d\tau'\right)S^{\dagger}$,
we have
\begin{align*}
\int_{0}^{\tau}\left\{ \int_{0}^{\tau'}\mathcal{H}_{L}^{10}\tilde{\rho}_{00}ds\right\} \mathcal{H}_{R}^{01}d\tau' & =\int_{0}^{\tau}2(L^{\dagger}L)_{11}^{-1}e^{\frac{\tau'}{2}(L^{\dagger}L)_{11}}\mathcal{T}e^{\int_{0}^{\tau'}i\epsilon\tilde{H}_{r}^{11}dr}H_{\tau'}^{10}\rho_{00}H_{\tau'}^{01}e^{\frac{\tau'}{2}(L^{\dagger}L)_{11}}\mathcal{T}e^{-\int_{0}^{\tau'}i\epsilon\tilde{H}_{s}^{11}ds}d\tau'+O(\epsilon)\\
 & =\int_{0}^{\tau}2(L^{\dagger}L)_{11}^{-1}e^{\frac{\tau'}{2}(L^{\dagger}L)_{11}}H_{\tau'}^{10}\rho_{00}H_{\tau'}^{01}e^{\frac{\tau'}{2}(L^{\dagger}L)_{11}}d\tau'+O(\epsilon)\\
\int_{0}^{\tau}\mathcal{H}_{L}^{10}\left[\int_{0}^{\tau'}\tilde{\rho}_{00}\mathcal{H}_{R}^{01}ds\right]d\tau' & =\int_{0}^{\tau}e^{\frac{\tau'}{2}(L^{\dagger}L)_{11}}H_{\tau'}^{10}\rho_{00}H_{\tau'}^{01}e^{\frac{\tau'}{2}(L^{\dagger}L)_{11}}2(L^{\dagger}L)_{11}^{-1}d\tau'+O(\epsilon)
\end{align*}
this implies that the sum of both terms above corresponds to an anticommutator, which can be written as a total derivative with an error of order $\epsilon$ as
\begin{align*}
\int_{0}^{\tau}\left\{ \int_{0}^{\tau'}\mathcal{H}_{L}^{10}\tilde{\rho}_{00}ds\right\} \mathcal{H}_{R}^{01}d\tau'+\int_{0}^{\tau}\mathcal{H}_{L}^{10}\left[\int_{0}^{\tau'}\tilde{\rho}_{00}\mathcal{H}_{R}^{01}ds\right]d\tau'  =\int_{0}^{\tau}\{e^{\frac{\tau'}{2}(L^{\dagger}L)_{11}}H_{\tau'}^{10}\rho_{00}H_{\tau'}^{01}e^{\frac{\tau'}{2}(L^{\dagger}L)_{11}},2(L^{\dagger}L)_{11}^{-1}\}d\tau'\\
  =2(L^{\dagger}L)_{11}^{-1}\left(\int_{0}^{\tau}\frac{d}{d\tau'}\left(e^{\frac{\tau'}{2}(L^{\dagger}L)_{11}}H_{\tau'}^{10}\rho_{00}H_{\tau'}^{01}e^{\frac{\tau'}{2}(L^{\dagger}L)_{11}}\right)d\tau'\right)2(L^{\dagger}L)_{11}^{-1}+O(\epsilon)\\
  =2(L^{\dagger}L)_{11}^{-1}\left(e^{\frac{\tau}{2}(L^{\dagger}L)_{11}}H_{\tau}^{10}\rho_{00}H_{\tau}^{01}e^{\frac{\tau}{2}(L^{\dagger}L)_{11}}\right)'2(L^{\dagger}L)_{11}^{-1}+O(\epsilon)
\end{align*}

so $\mathcal{I}$ becomes
\[
\mathcal{I}=4U(\tau,0)L_{01}(L^{\dagger}L)_{11}^{-1}H_{\tau}^{10}\rho_{00}H_{\tau}^{01}(L^{\dagger}L)_{11}^{-1}L_{10}^{\dagger}U^{\dagger}(\tau,0)+O(\epsilon).
\]
Putting all together back in Eq. \ref{eq:rho00} gives
\begin{align}\label{eq:Lindblad}
\frac{d}{d\tau}\rho_{00}+i\epsilon[H_{\tau}^{00},\rho_{00}] & =\epsilon^{2}\left(4L_{01}(L^{\dagger}L)_{11}^{-1}H_{\tau}^{10}\rho_{00}H_{\tau}^{01}(L^{\dagger}L)_{11}^{-1}L_{10}^{\dagger}-\frac{1}{2}\left\{ H_{\tau}^{01}4(L^{\dagger}L)_{11}^{-1}H_{\tau}^{10},\rho_{00}\right\} \right)+O(\epsilon^{3})
\end{align}
this equation has the Lindblad structure
\begin{align*}
\frac{d}{d\tau}\rho_{00}=-i\epsilon[H_{\tau}^{00},\rho_{00}]+\epsilon^{2}\left(\ell_{\tau}\rho_{00}\ell_{\tau}^{\dagger}-\frac{1}{2}\left\{ \ell_{\tau}^{\dagger}\ell_{\tau},\rho_{00}\right\} \right)+O(\epsilon^{3})
\end{align*}
with $\ell_{\tau}=2L_{01}(L^{\dagger}L)_{11}^{-1}H_{\tau}^{10}$.
This proves our main result. 

\subsection{Gauge invariance of the DS}\label{app:Gauge}

Let's define the non-Abelian gauge field $A_{\mu}^{0}=iP_{0}(\partial_{\mu}\mathcal{U})\mathcal{U}^{\dagger}P_{0}$,
where $x^{\mu}=x^{\mu}(\tau)$ are the parameters of $L_{\tau}$
that vary over time. The effective Hamiltonian $H_{0}$ is then given by
$H_{0}={\gamma T}\sum_{\mu}A_{\mu}^{0}\frac{dx^{\mu}}{d{\tau}}$. One can rotate states within the dark space with
a unitary matrix $\omega$, such that 
\begin{align}
L_{\tau}=\mathcal{U}_{\tau}^{\dagger}L\mathcal{U}_{\tau}=\mathcal{U}_{\tau}^{\dagger}\omega^{\dagger}(\omega L\omega^{\dagger})\omega \mathcal{U}_{\tau}.
\end{align}
The rotated operator $L^\omega=\omega L\omega^{\dagger}$ annihilates the original
dark space iff $[\omega,P_{0}]=0$. Using this new basis,
one obtains another Hamiltonian $H_{0}^{\omega}={\gamma T}\sum_{\mu}(A_{\mu}^{0})^{\omega}\frac{dx^{\mu}}{d{\tau}}$, where 
\begin{align}
(A_{\mu}^{0})^{\omega} & =\omega^{0}A^0_{\mu}\omega^{\dagger0}+i(\partial_{\mu}\omega^{0})\omega^{\dagger0},
\end{align}
and $\omega^{0}=P_{0}\omega P_{0}$. This corresponds to the non-Abelian
gauge transformation of the gauge field $A_{\mu}^{0}$. 
Such gauge transformation is inconsequential for the non-Abelian Berry phase defined on a closed loop in the parameter space.
As we show below, this transformation  affects the effective quantum jump operators.  The latter transform covariantly, implying that any quantity defined as a trace of products of the jump operators is gauge invariant.
Indeed, the gauge transformation above induces a change of the effective quantum
jump operators as 
\begin{align*}
&X^{\omega}_\tau=  \int\limits_{0}^{\tau}ds\, e^{\frac{1}{2}L^{\omega\dagger}L^{\omega}(s-\tau)}\, (1-P_{0})H^{\omega}_sP_{0}\\
&=\!  \int\limits_{0}^{\tau}\!ds\, \omega\, e^{\frac{1}{2}L^{\dagger}L(s-\tau)}\omega^{\dagger}(1-P_{0})\left(\omega H\omega^{\dagger}+i\gamma T(\partial_{s}\omega)\omega^{\dagger}\right)P_{0}\\
&=  \omega X_\tau\omega^{\dagger},
\end{align*}
where we have discarded terms of order $1/(\gamma T)$ and used that
$[\omega_{\tau},P_{0}]=0$ implies $(1-P_{0})(\partial_{\mu}\omega)\omega^{\dagger}P_{0}=0.$
From here we have $\ell{}_{\tau}^{\omega}=P_{0}L^{\omega}\omega X_\tau\omega^{\dagger}P_{0}=\omega\ell_{\tau}\omega^{\dagger}.$
This implies that under gauge transformations, the effective quantum jump operator transforms covariantly, as expected.

\end{document}